\documentclass[a4paper,11pt]{article}

\usepackage{jheppub} 

\usepackage[T1]{fontenc} 

\usepackage{graphicx}
\usepackage{float}

\usepackage{subcaption}

\title{\boldmath Conformal spacelike-timelike correspondence in QCD}

\author[a]{A. H. Mueller}

\affiliation[a]{Physics Department, Columbia University, New York, NY 10027}

\abstract{
This paper is a study of a spacelike-timelike conformal correspondence in QCD. When the times at vertices are fixed in an $A_+=0$ gauge calculation the distribution of gluons in a highly virtual decay have an exact correspondence with the gluons in the lightcone wavefunction of a high energy dipole with the identification of angles in the timelike case and transverse coordinates in the lightcone wavefunction. Divergences show up when the time integrals are done. A procedure for dropping these divergences, analogous to the Gell-Mann Low procedure in QED, allows one to define a conformal QCD, at least through NLO. Possible uses of such a conformal QCD are discussed.
}

\begin{document} 
\maketitle
\flushbottom

\section{Introduction}

Over the past fifteen years or so it has become increasingly clear that there are nontrivial relations between the distribution of particles in the decay of a highly timelike current and properties of high energy scattering processes. The first hint of such relations was the fact that the BMS equation\cite{1}, an equation which was developed to describe nonglobal properties of jet decays\cite{2}, is essentially identical to the BK equation\cite{3,4}, an equation describing high energy scattering. Shortly after the appearance of the BMS equation it was discovered\cite{5,6,7} that in certain kinematic regions of a jet decay the number of produced heavy quarks, or minijets, is given by the BFKL equation\cite{8,9}, an equation long used to describe high energy hard scattering away from the unitarity limit.
 
 The relationship between jet decays and high energy scattering became more interesting when Hofman and Maldacena\cite{10} and Hatta\cite{11} recognized that in the AdS/CFT correspondence the angular distribution of energy and charge in the decay of a highly virtual current is directly related to the transverse coordinate distribution of these same quantities in a high energy \textit{hadron}. Hatta\cite{11} then exhibited a stereographic projection relating the angular distribution of these quantities in jet decays to their transverse coordinate distributions in a high energy hadron, thus making the conformal relationship more explicit.
 
 However, this is all a bit mysterious. Jet decays and the corresponding distribution of energy and particle densities are physical while the wave function of a high energy hadron is gauge and quantization dependent. To avoid this issue one could simply interpret the spacelike-timelike equivalence as one of evolution, In \cite{11} the equivalence of BMS evolution (timelike) to BK evolution (spacelike) was demonstrated while in \cite{12} the double logarithmic resummations necessary to tame the NLO kernels in BMS and BK evolution were shown to be related. However, the correspondence appears to be stronger than just an equivalence of evolutions.
 
 In this paper we compare the distribution of particles in the decay of a timelike current into a quark-antiquark pair, along with an arbitrary number of gluons, with the distribution of partons(gluons) in the lightcone wavefunction of a high energy dipole\cite{13} and find a one to one correspondence. More precisely, in the decay of a timelike current we suppose that the quark and antiquark, initially produced by the current, have longitudinal momenta much greater than that of the soft gluons subsequently emitted, and we fix $\theta_{ab}$ to be the angle between the quark and the antiquark in a highly boosted frame where $\theta_{ab} \ll 1$. (For simplicity we suppress additional quark-antiquark production.) On the \textit{hadron} side we suppose an initial quark-antiquark dipole with transverse coordinate separation $x_{\bot ab}$ and we only consider gluons in the $A_+=0$ lightcone wavefunction whose longitudinal momenta are much less than that of the parent quark and antiquark dipole. Further, we suppose the quark and antiquark longitudinal momenta are identical in the decay and in the high energy dipole wavefunction. In the decay into soft gluons we do not suppose any strong ordering among the longitudinal momenta of the gluons, but later we shall only explicitly consider evolutions at NLO due to subtleties of coupling renormalization. The requirement that the gluon momenta be soft compared to the parent quark-antiquark pair is, we believe, an essential assumption. If a gluon had longitudinal momentum comparable to that of the quark-antiquark dipole(spacelike case), then that gluon emission would be sensitive to how the parent dipole was created and we believe that is beyond the correspondence we are considering.  
 
At a given order of perturbation theory we observe a graph by graph equivalence for a timelike decay probability of $\gamma^*(Q)\rightarrow q(p_a)+\bar{q}(p_b)+g_1(\underline{\theta}_1, k_{1+})+\cdots+g_m(\underline{\theta}_m, k_{m+})$ and the square of the dipole  wavefunction of $\psi(p_a, p_b, g_1(\underline{\theta}_1, k_{1+}), \cdots, g_m(\underline{\theta}_m, k_{m+}))$ when we identify $\underline{\theta}_i=\sqrt{2}\underline{k}_i/k_{i+}$ with $\underline{x}_i$. The timelike and spacelike quantities are written as integrands over which integrations over the times at all the vertices present in a given graph are to be done. The integrands of the two processes, with the $\underline{\theta}_i \leftrightarrow \underline{x}_i$ identification, are identical with no restriction on the gluon momenta except that the longitudinal momentum of every gluon must be small compared to the parent quark and antiquark momentum.

However, there are divergences when the time integrations are done. In some cases, when a time $t_i$ in the tinelike proess goes to infinity, corresponding to a time $t_i$ in the spacelike process going to zero, there are other graphs which cancel these divergences. These are "real-virtual" cancellations. (In the timelike case the cancellation will happen when one measures a jet rather than an individual particle, while in the spacelike case the cancellation will happen when the real and virtual configurations are not distinguished by a scattering.) These cancellations are always of collinear singularities in the timelike case and ultraviolet singularities in the spacelike case.

Other corresponding singularities do not cancel. They are ultraviolet singularities in both spacelike and timelike cases and represent the necessity of coupling renormalization in QCD. The introduction of the QCD $\Lambda$-parameter breaks the conformal invariance and with it the spacelike-timelike correspondence. In section \ref{sec:4} we suggest a precise way of removing the coupling divergences, much like that originally done by Gell-Mann and Low\cite{14} for QED, occurring only in self-energy graphs in our $A_+=0$ gauge dynamics. This removal does not introduce any new scale and leaves a "conformal QCD" and a correspondence between spacelike and timelike processes. Howevver, we have only been able to demonstrate this subtraction throughNLO in soft emissions.

One of the most ambitious, and interesting, programs using the spacelike-timelike correspondence has been that of Caron-Huot\cite{15} who showed that in $N=4$ SYM the NLO kernel for BK evolution\cite{16,17,18} could be obtained purely from the evaluation of decays. His procedure does not work when the $\beta$-function is not zero. However, in this case one should be able to evaluate the timlike process with certain (see section \ref{sec:4}) self-energy graphs in $A_+=0$ gauge removed, translate that to the contribution to the NLO BK kernel and then add the self-energy contributions back in with the appropriate renormalization in the spacelike process.

\section{An example and its generalization}
\label{sec:2}

\begin{figure}[H]
\begin{subfigure}[b]{1.0\textwidth}
	\centering
	\includegraphics[]{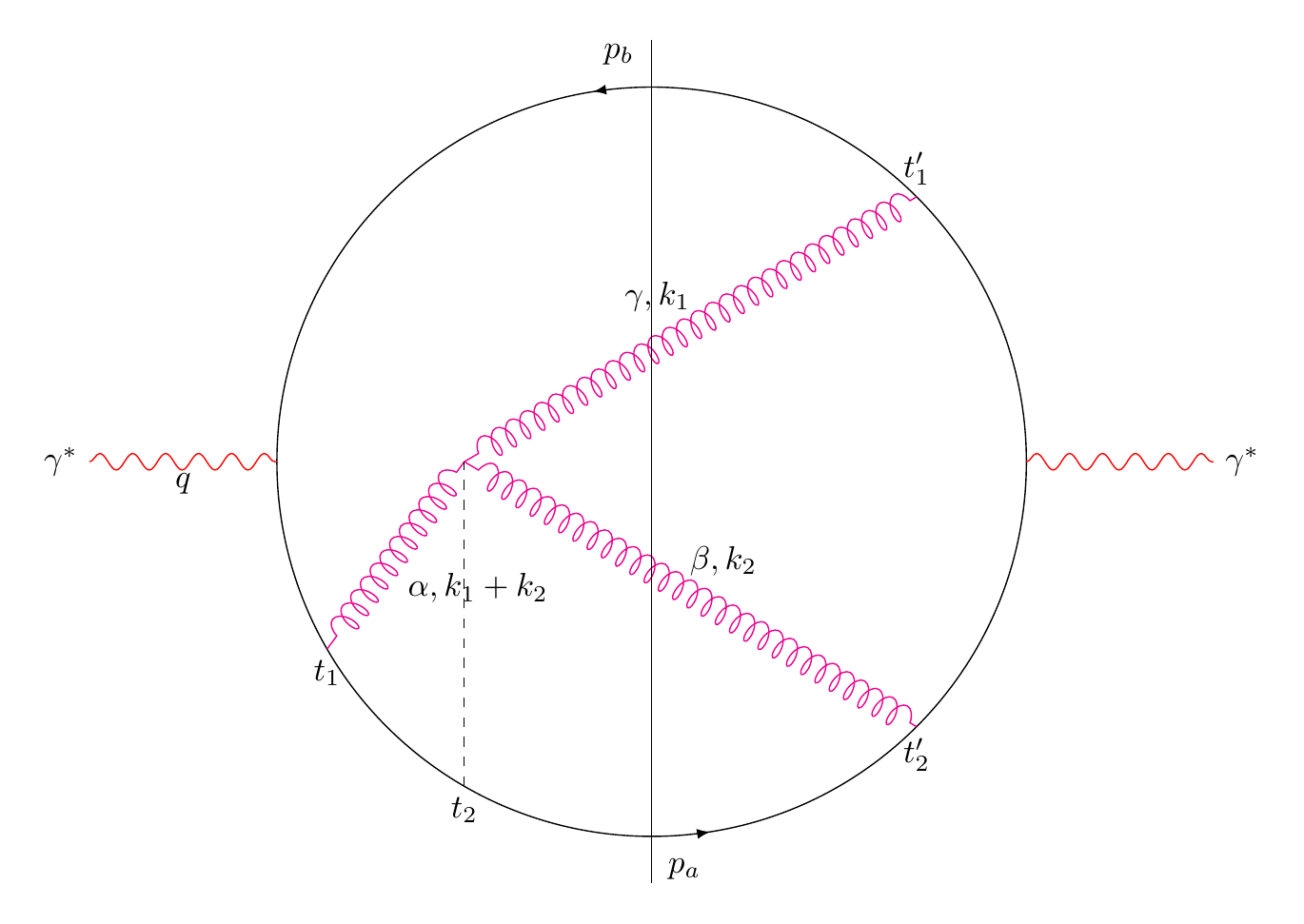}
	\caption{fig:1a.}\label{fig:1a}
\end{subfigure}

\begin{subfigure}[b]{1.0\textwidth}
	\centering
	\includegraphics[]{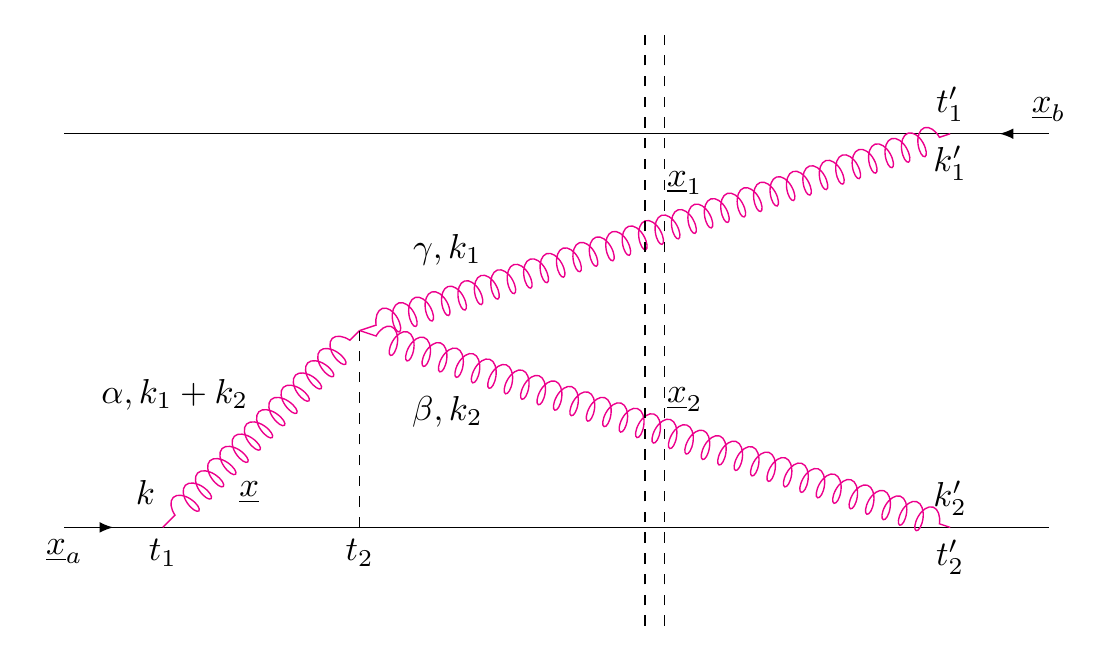}
	\caption{fig:1b.}\label{fig:1b}
\end{subfigure}
\caption{Sample graphs illustrating the conformal correspondence.}\label{fig:1}
\end{figure}

We start with a nontrivial example of a graph having a three gluon vertex as well as couplings to the parent quarks in which the conformal correspondence of the graph as part of the decay of a timelike photon to the graph as part of the lightcone wavefunction of a high energy dipole will be exhibited.The graphs are illustrated in Figure \ref{fig:1}. We work in a frame where the timelike virtuality $Q$ of the photon, $q$, in Figure \ref{fig:1a} obeys $Q/q_+ \ll 1$ so that the angle $\theta_{ab}$ between the quark $a$ and the antiquark $b$ is very small. For the $A_+=0$ lightcone wavefunction illustrated in Figure \ref{fig:1b} the lines will be labelled by a transverse coordinate and a longitudinal momentum, although to begin we write the wavefunction only in terms of gluon momenta. The correspondence will relate the decay rate($a$), at given time values $t_i$ at the vertices and fixed $(\underline{t}_i, k_{i+})$ on each of the lines to the square of the lightcone wavefunction($b$), also for fixed times at each of the vertices but with the corresponding lines labelled by $(\underline{x}_i,k_{i+})$. In the correspondence $\underline{x}_i$ and $\underline{k}_i$ are related by
\begin{equation}
\underline{\theta}_i=\frac{\sqrt{2}\underline{k}_i}{k_{i+}}\leftrightarrow\underline{x}_i.	
\end{equation}
We begin by writing the graph, corresponding to a decay, of Figure \ref{fig:1a} in detail. Then we shall write the corresponding graph for the square of the lightcone wavefunction, shown in Figure \ref{fig:1b}, and observe that they are the same. We always assume that the fermion lines, $a$ and $b$, have a much larger longitudinal momentum than the gluon lines but there will be no assumed ordering as to the relative magnitude of $k_{1+}$ and $k_{2+}$.

\subsection{The decay graph of figure \ref{fig:1a}}

The decay rate of the virtual photon without radiative correction is $W_0$. If $W$ is the rate with radiative corrections, then we are going to write an expression for $w=W/W_0$ as
\begin{equation}
\label{eq:2}
	w=\frac{-ig^4}{(2\pi)^4}\frac{N_c^2-1}{4}\int A_L(\underline{\theta}_1,\underline{\theta}_2,k_{1+},k_{2+})\cdot A_R^*(\underline{\theta}_1,\underline{\theta}_2,k_{1+},k_{2+})\frac{dk_{1+}}{2k_{1+}}\frac{dk_{2+}}{2k_{2+}}\frac{k_{1+}^2}{2}d^2\theta_1\frac{k_{2+}^2}{2}d^2\theta_2
\end{equation}
for the graph of figure \ref{fig:1a}. We shall then identify $A_L$ and $A_R$ with corresponding expressions for the graph of figure \ref{fig:1b} with the time integrations fixed in each expression. Further write the part to the left of the cut, $A_L$, as
\begin{equation}
\label{eq:3}
	A_L=a_Lb_L,
\end{equation}
with a similar separation for $A_R$, where $a_L$ includes exponential factor and time integrations while vertex factors are included in $b_L$. Then
\begin{equation}
	a_L=\int_0^\infty dt_1\int_{t_1}^\infty dt_2 e^{i\Delta E_1t_1+i\Delta E_2t_2}
\end{equation}
where
\begin{equation}
	\Delta E_1=\frac{(\underline{k}_1+\underline{k}_2)^2}{2(k_1+k_2)_+}+\frac{\underline{p}_a^2}{2p_{a+}}-\frac{(\underline{p}_{a}+\underline{k}_1+\underline{k}_2)^2}{2(p_a+k_1+k_2)_+}
\end{equation}
and
\begin{equation}
	\Delta E_2=\frac{\underline{k}_1^2}{2k_{1+}}+\frac{\underline{k}_2^2}{2k_{2+}}-\frac{(\underline{k}_1+\underline{k}_2)^2}{2(k_1+k_2)_+}
\end{equation}
It is straightforward to get
\begin{equation}
	\Delta E_1=\frac{(k_1+k_2)_+}{4}(\underline{\theta}-\underline{\theta}_a)^2
\end{equation}
\begin{equation}
	\Delta E_2=\frac{k_{1+}k_{2+}}{4(k_1+k_2)_+}(\underline{\theta}_1-\underline{\theta}_2)^2
\end{equation}
where
\begin{equation}
	\underline{\theta}=\frac{1}{(k_1+k_2)_+}[k_{1+}\underline{\theta}_1+k_{2+}\underline{\theta}_2].
\end{equation}
Thus
\begin{equation}
	a_L=\int_0^\infty dt_1\int_{t_1}^\infty dt_2 \exp\left\{\frac{i}{4}\left[(k_1+k_2)_+(\underline{\theta}-\underline{\theta}_a)^2t_1+\frac{k_{1+}k_{2+}}{(k_1+k_2)_+}(\underline{\theta}_1-\underline{\theta}_2)^2 t_2\right]\right\}
\end{equation}
Similarly
\begin{equation}
	a_R^*=\int_0^\infty dt_1'\int_{0}^\infty dt_2' \exp\left\{\frac{-i}{4}\left[k_{1+}(\underline{\theta}_1-\underline{\theta}_b)^2t_1'+k_{2+}(\underline{\theta}_2-\underline{\theta}_a)^2 t_2' \right]\right\}.
\end{equation}

Now turn to the vertex factors, the $b$ term in (\ref{eq:3}). In the amplitude of the graph of \ref{fig:1a} there is a vertex at $t_1$ and a three-gluon vertex at $t_2$. Call $b_L=P_{1L}P_{2L}$ where the vertex $P_{1L}$ is given by
\begin{equation}
\label{eq:12}
	P_{1L}=\frac{\bar{u}(p_a)\gamma\cdot\epsilon^\lambda}{\sqrt{2p_{a+}}}\frac{u(p_a+k_1+k_2)}{\sqrt{2(p_a+k_1+k_2)_+}}\simeq \frac{\underline{\epsilon}^\lambda\cdot(\underline{k}_1+\underline{k}_2)}{(k_1+k_2)_+}-\frac{\underline{p}_a\cdot\underline{\epsilon}^\lambda}{p_{a+}}
\end{equation}
or
\begin{equation}
\label{eq:13}
	P_{1L}=\frac{1}{\sqrt{2}}[\underline{\theta}-\underline{\theta}_a]\cdot\underline{\epsilon}^\lambda
\end{equation}
In reaching (\ref{eq:13}) we have assumed that $p_{a+}\gg k_{1+},k_{2+}$ but we suppose that $\underline{\theta}$, $\underline{\theta}_1$ and $\underline{\theta}_2$ may all be of comparable magnitude.

The three gluon vertex, $P_{2L}$, is given by
\begin{equation}
	P_{2L}=\epsilon_\alpha^\lambda\epsilon_\gamma^{\lambda_1}\epsilon_\beta^{\lambda_2}\left[-g_{\alpha\gamma}(2k_1+k_2)_\beta+g_{\alpha\beta}(2k_2+k_1)_\gamma-g_{\alpha\gamma}(k_2-k_1)_\alpha\right]
\end{equation}
or
\begin{equation}
\label{eq:15}
	P_{2L}=\sqrt{2}(\underline{\theta}_2-\underline{\theta}_1)\cdot\left[k_{1+}\underline{\epsilon}^{\lambda_2}(\underline{\epsilon}^{\lambda}\cdot\underline{\epsilon}^{\lambda_1})+k_{2+}\underline{\epsilon}^{\lambda_1}(\underline{\epsilon}^{\lambda}\cdot\underline{\epsilon}^{\lambda_2})-\frac{k_{1+}k_{2+}}{(k_1+k_2)_+} \underline{\epsilon}^{\lambda}(\underline{\epsilon}^{\lambda_1}\cdot\underline{\epsilon}^{\lambda_2})\right]
\end{equation}
In (\ref{eq:12})-(\ref{eq:15}) we imagine using real polarization vectors in order to avoid a proliferation of complex conjugate symbols. An abbreviated notation is being used where $
\epsilon^{\lambda_i}=\epsilon^{\lambda_i}(k_i)$ and $\epsilon^\lambda=\epsilon^\lambda (k_1+k_2)$. $b_L$ is obtained as
\begin{equation}
\label{eq:16}
	b_L=\sum_{\lambda}P_{1L}P_{2L}=(\theta-\theta_a)_i(\theta_2-\theta_1)_j\left[k_{1+}\epsilon_i^{\lambda_1}\epsilon_j^{\lambda_2}+k_{2+}\epsilon_i^{\lambda_2}\epsilon_j^{\lambda_1}-\delta_{ij}\frac{k_{1+}k_{2+}}{(k_1+k_2)_+}\underline{\epsilon}^{\lambda_1}\cdot\underline{\epsilon}^{\lambda_2}\right].
\end{equation}
$b_R$ is easily found to be
\begin{equation}
\label{eq:17}
	b_R=\frac{1}{2}(\underline{\theta}_2-\underline{\theta}_a)\cdot\underline{\epsilon}^{\lambda_2}(\underline{\theta}_1-\underline{\theta}_b)\cdot\underline{\epsilon}^{\lambda_1}.
\end{equation}

Thus the integrand in (\ref{eq:2}) is given by
\begin{align}
\label{eq:18}
		A_LA_R^*=&\int_0^\infty dt_1 \int_{t_1}^\infty dt_2 \int_0^\infty dt_1' \int_0^\infty dt_2' \nonumber \\
	&\times\exp\left\{\frac{i}{4}\left[
	(k_1+k_2)_+ (\theta-\theta_a)^2 t_1-k_{1+}(\theta_1-\theta_b)^2 t_1'-k_{2+}(\theta_2-\theta_a)^2 t_2' + \frac{k_{1+}k_{2+}}{(k_1+k_2)_+}(\theta_1-\theta_2)^2 t_2
	\right]\right\} \nonumber \\
	&\times\frac{1}{2}\sum_{\lambda_1,\lambda_2}b_L\cdot b_R.
\end{align}
Equation (\ref{eq:18}) with $b_L$ and $b_R$ given by (\ref{eq:16}) and (\ref{eq:17}) respectively is a convenient form for the decay to compare to the high energy dipole wavefunction which we turn to next.

\subsection{The high energy wave function graph of figure \ref{fig:1b}}
Our goal is to express the square of the high energy wavefunction contained in figure \ref{fig:1b} in terms of an integration over coordinates $d^2\underline{x}_1d^2\underline{x}_2$ and to identify the integrand with (\ref{eq:18}). We begin in momentum space and write the vertices as
\begin{equation}
\label{eq:19}
	V_1=e^{\frac{i\underline{k}^2t_1}{2k_+}}\frac{\underline{\epsilon}^\lambda\cdot\underline{k}}{k_+}e^{-i\underline{k}\cdot\underline{x}_a}
\end{equation}
\begin{equation}
	V_1'^*=e^{-\frac{i(\underline{k}_1')^2t_1'}{2k_{1+}}}\frac{\underline{\epsilon}^{\lambda_1}\cdot\underline{k}_1'}{k_{1+}}e^{i\underline{k}_1'\cdot\underline{x}_b}
\end{equation}
\begin{equation}
\label{eq:21}
	V_2'^*=e^{-\frac{i(\underline{k}_2')^2t_2'}{2k_{2+}}}\frac{\underline{\epsilon}^{\lambda_2}\cdot\underline{k}_2'}{k_{2+}}e^{i\underline{k}_2'\cdot\underline{x}_a}
\end{equation}
\begin{equation}
\label{eq:22}
	V_2=e^{i\left[\frac{\underline{k}_1^2}{2k_{1+}}+\frac{\underline{k}_2^2}{2k_{2+}}-\frac{(\underline{k}_1+\underline{k}_2)^2}{2k_{+}}\right]t_2}
	\epsilon_\alpha^\lambda \epsilon_\gamma^{\lambda_1} \epsilon_\beta^{\lambda_2}
	\left[-g_{\alpha\gamma}(2k_1+k_2)_\beta+g_{\alpha\beta}(2k_2+k_1)_\gamma-g_{\gamma\beta}(k_2-k_1)_\alpha\right]
\end{equation}
where $k_+=k_{1+}+k_{2+}$, $k_{1+}=k_{1+}'$, $k_{2+}=k_{2+}'$ but where, 
for the moment, we do not take $\underline{k}_1$ and $\underline{k}_1'$ or $\underline{k}_2$ and $\underline{k}_2'$ to be equal. Instead we put a coordinate on each line with phase factors which, after the coordinates are integrated, given transverse momentum conservation. Then in addition to the factors above we include the factors $L_1$, $L_2$, $L$ where
\begin{equation}
\label{eq:23}
	L_1=\frac{d^2x_1}{(2\pi)^2}e^{i(\underline{k}_1-\underline{k}_1')\cdot\underline{x}_1}
\end{equation}
\begin{equation}
\label{eq:24}
	L_2=\frac{d^2x_2}{(2\pi)^2}e^{i(\underline{k}_2-\underline{k}_2')\cdot\underline{x}_2}
\end{equation}
\begin{equation}
\label{eq:25}
	L=\frac{d^2x}{(2\pi)^2}e^{i(\underline{k}-\underline{k}_1-\underline{k}_2)\cdot\underline{x}}.
\end{equation}
Clearly the integrations over $\underline{x}$, $\underline{x}_1$ and $\underline{x}_2$ give transverse momentum conservation.

In analogy with the previous section, we group the factors together as
\begin{equation}
\label{eq:26}
	\bar{A}_L=\int_{-\infty}^0 dt_1 \int_{t_1}^0 dt_2V_1V_2Ld^2kd^2k_1d^2k_2.
\end{equation}
The various $k$-integrals in (\ref{eq:26}) are easily done
\begin{equation}
\label{eq:27}
	\tilde{V}_1=\int d^2k V_1e^{i\underline{k}\cdot\underline{x}}=\frac{-2\pi i \underline{\epsilon}^\lambda\cdot(\underline{x}-\underline{x}_a)(k_1+k_2)_+}
	{t_1^2}
	e^{\frac{-i(\underline{x}-\underline{x}_a)^2(k_1+k_2)_+}{2t_1}}
\end{equation}
\begin{equation}
\label{eq:28}
	\tilde{V}_2=\int d^2k_1d^2k_2 V_2 e^{-i(\underline{k}_1+\underline{k}_2)\cdot\underline{x}+i\underline{k}_1\cdot\underline{x}_1+i\underline{k}_2\cdot \underline{x}_2}.
\end{equation}
Using (\ref{eq:22}) one finds
\begin{align}
\label{eq:29}
	\tilde{V}_2=&\frac{-2i(\underline{x}_2-\underline{x}_1)}{t_2^2}\cdot\left[k_{1+}\underline{\epsilon}^{\lambda_2}(\underline{\epsilon}^{\lambda}\cdot\underline{\epsilon}^{\lambda_1})+k_{2+}\underline{\epsilon}^{\lambda_1}(\underline{\epsilon}^{\lambda}\cdot\underline{\epsilon}^{\lambda_2})-\underline{\epsilon}^{\lambda}(\underline{\epsilon}^{\lambda_2}\cdot\underline{\epsilon}^{\lambda_1})\frac{k_{1+}k_{2+}}{(k_1+k_2)_+}
	\right] \nonumber\\
	&\times (2\pi)^3\frac{k_{1+}k_{2+}}{(k_1+k_2)_+}\delta\left(\underline{x}-\frac{k_{1+}}{(k_1+k_2)_+}\underline{x}_1-\frac{k_{2+}}{(k_1+k_2)_+}\underline{x}_2\right)
	e^{\frac{-i(\underline{x}_1-\underline{x}_2)^2k_{1+}k_{2+}}{2t_2(k_1+k_2)_+}}
\end{align}
Using (\ref{eq:27}) and (\ref{eq:29}) in (\ref{eq:26}) along with $\tau_1=\frac{2}{t_1}$, $\tau_1=-\frac{2}{t_2}$ gives
\begin{align}
\label{eq:30}
		\bar{A}_L=&-\int_0^\infty d\tau_1 \int_{\tau_1}^\infty d\tau_2
	e^{i(\underline{x}-\underline{x}_a)^2(k_1+k_2)_+\frac{\tau_1}{4}+\frac{i(\underline{x}_1-\underline{x}_2)^2\underline{k}_{1+}\underline{k}_{2+}}{4(k_1+k_2)_+}\tau_2}2\pi^2d^2x(k_{1+}k_{2+}) \nonumber \\
	&\times (x-x_a)_i(x_2-x_1)_j\left[k_{1+}\epsilon_i^{\lambda_1}\epsilon_j^{\lambda_2}+k_{2+}\epsilon_j^{\lambda_1}\epsilon_i^{\lambda_2}-\delta_{ij}\frac{k_{1+}k_{2+}}{(k_1+k_2)_+}\underline{\epsilon}^{\lambda_1}\cdot\underline{\epsilon}^{\lambda_2}\right] \nonumber \\
	&\times\delta\left(\underline{x}-\frac{k_{1+}}{(k_1+k_2)_+}\underline{x}_1-\frac{k_{2+}}{(k_1+k_2)_+}\underline{x}_2\right).
\end{align}
Similarly $\bar{A}_R^*$ defined by
\begin{equation}
	\bar{A}_R^*d^2x_1d^2x_2=\int_{-\infty}^0 dt_1' \int_{-\infty}^0 dt_2' V_1'^* V_2'^* L_1 L_2 d^2k_1' d^2k_2'
\end{equation}
is easily evaluated to be
\begin{align}
\label{eq:32}
	\bar{A}_R^*d^2x_1d^2x_2=&-\int_0^\infty d\tau_1' \int_0^\infty d\tau_2' e^{-i(\underline{x}_1-\underline{x}_b)^2k_{1+}\frac{\tau_1'}{4}-i(\underline{x}_2-\underline{x}_a)^2k_{1+}\frac{\tau_2'}{4}}\frac{d^2x_1d^2x_2k_{1+}k_{2+}}{4(2\pi)^2} \nonumber \\
	&\times \underline{\epsilon}^{\lambda_1}\cdot(\underline{x}_1-\underline{x}_b)\underline{\epsilon}^{\lambda_2}\cdot(\underline{x}_2-\underline{x}_a).
\end{align}
Multiplying (\ref{eq:30}) and (\ref{eq:32}) and doing the sum over $\lambda_1$,$\lambda_2$ gives, in analogy with (\ref{eq:18}),
\begin{align}
\label{eq:33}
		\bar{A}_L\bar{A}_R^*&d^2x_1d^2x_2=\int_0^\infty d\tau_1 \int_{\tau_1}^\infty d\tau_2 \int_0^\infty d\tau_1' \int_0^\infty d\tau_2' \nonumber \\
	&\times\exp\left\{\frac{i}{4}\left[
	(k_1+k_2)_+ (\underline{x}-\underline{x}_a)^2 \tau_1-k_{1+}(\underline{x}_1-\underline{x}_b)^2 \tau_1'-k_{2+}(\underline{x}_2-\underline{x}_a)^2 \tau_2' + \frac{k_{1+}k_{2+}}{(k_1+k_2)_+}(\underline{x}_1-\underline{x}_2)^2 \tau_2
	\right]\right\} \nonumber \\
	&\times\frac{1}{8}\sum_{\lambda_1,\lambda_2}\bar{b}_L\cdot \bar{b}_R(k_{1+}k_{2+})^2d^2x_1d^2x_2
\end{align}
where $\bar{b}_L$ and $\bar{b}_R$ are identical the $b_L$ and $b_R$, in (\ref{eq:16}) and (\ref{eq:17}), with the replacement $\underline{\theta}_i,\underline{\theta}\rightarrow \underline{x}_i,\underline{x}$. To make the correspondence precise write $w$ in (\ref{eq:2}) as
\begin{equation}
\label{eq:34}
	w=\frac{-ig^4(N_c^2-1)}{4(2\pi)^4}\int_0^\infty dt_1 \int_{t_1}^\infty dt_2 \int_0^\infty dt_1' \int_0^\infty dt_2'
	I \frac{dk_{1+}}{2k_{1+}}\frac{dk_{2+}}{2k_{2+}}\frac{k_{1+}^2}{2}d^2\theta_1\frac{k_{2+}^2}{2}d^2\theta_2
\end{equation}
and write the amount of probability that graph \ref{fig:1b} contributes to the square of the dipole wavefunction as
\begin{equation}
\label{eq:35}
	\bar{w}=\frac{-ig^4(N_c^2-1)}{4(2\pi)^4}\int_0^\infty d\tau_1 \int_{\tau_1}^\infty d\tau_2 \int_0^\infty d\tau_1' \int_0^\infty d\tau_2'
	\bar{I} \frac{dk_{1+}}{2k_{1+}}\frac{dk_{2+}}{2k_{2+}}\frac{k_{1+}^2}{2}d^2x_1\frac{k_{2+}^2}{2}d^2x_2
\end{equation}
then $I=\bar{I}$ when the $t_i$, $\underline{\theta}_i$ variables of the $I$ are identified with the $\tau_i$, $\underline{x}_i$ variables of $\bar{I}$.

Although we are identifying variables with different dimension in the correspondence we note that both $w$ and $\bar{w}$ are dimensionlesss so that one could always introduce a (fictitious) dimensional parameter to scale $x_i$, $t_i$ and $\tau_i$ to dimensionless varaibles.

In dealing with the graphs of figure \ref{fig:1} we have separated the graphs into vertices and lines, as for example in (\ref{eq:19})-(\ref{eq:21}) and (\ref{eq:23})-(\ref{eq:25}). It should be clear that for any graph built out of three-gluon vertices and causal propagation the procedure we have used here will work and lead to a correspondence between the probability of a given configuration of gluons appearing in the decay of a timelike photon and the probability that the corresponding gluons appear in the square of the lightcone wavefunction. It is straightforward to see that the correspondence continues to be valid when four-gluon vertices and instantaneous propagation is included, but we omit the details here for simplicity.

Our result might seem to be too strong. After all, we expect the decay-wavefuntion correspondence to reflect conformal symmetry and it is known that running coupling corrections will break conformal symmetry. So how does the breaking of conformal symmetry come into our discussion? The correspondence identifying $I$ in (\ref{eq:34}) with $\bar{I}$ in (\ref{eq:35}), once $t_i$, $\theta_i$ variables in $I$ have been changed to $\tau_i$, $x_i$ variables to get $\bar{I}$ is for fixed times. We believe this correspondence to be exact. However, the integrations over $dt_i$ and $d\tau_i$ have divergences when two times approach each other. In some circumstances these divergences can be removed simply by considering a more appropriate "jet" variable. In other circumstances these divergences must be removed by renormalization. Renormalization requires introducing a scale which breaks the conformal symmetry and that breaking corresponds to the running of the coupling in QCD. The graphs we have considered in this section have no divergences when the time integrations are done and so the correspondence survives time integration. In the next section of this paper we consider graphs which include running coupling effects.

\section{Graphs with running coupling divergences}
\label{sec:3}
We now turn to graphs having running coupling corrections, in particular the two graphs shown in figure \ref{fig:2}. We begin with graph \ref{fig:2b}. For fixed $t_1$, $t_2$, $t_3$, $t_4$ it is straightforward to write the graph as
\begin{align}
\label{eq:36}
	\bar{w}=&\frac{-g^4(N_c^2-1)}{2(2\pi)^4}\int \frac{dk_{1+}}{2k_{1+}} \frac{dk_{2+}}{2k_{2+}} \frac{d^2kd^2k'}{(2k_+)^2}
	V_1^\lambda (V_4^{\lambda'})^* V_2^{\lambda \lambda_1 \lambda_2}(V_3^{\lambda'\lambda_1\lambda_2})^* \nonumber \\
	&\times LL_1L_2(L')^* \prod_{i=1}^2 d^2k_id^2k'_i dt_1 dt_2 dt_3 dt_4
\end{align}
where $L$, $L_1$ and $L_2$ are as in (\ref{eq:23})-(\ref{eq:25}) while $L'$ is the same as (\ref{eq:25}) after the replacement $\underline{k},\underline{k}_i,\underline{x}\leftrightarrow \underline{k}',\underline{k}'_i,\underline{x}'$. A sum over all $\lambda$'s is understood in (\ref{eq:36}), while $V_2$ is as in (\ref{eq:22}) and $V_3$ is obtained from $V_2$ by the replacements $t_2,\underline{k}_i,\lambda \leftrightarrow t_3,\underline{k}'_i,\lambda'$. The limits on the $dt_i$-integrations will be given later.

Call
\begin{equation}
\label{eq:37}
	I^{\lambda\lambda'}=\int \prod_{i=1}^2 d^2k_i d^2k'_i e^{
	i\underline{k}_1\cdot(\underline{x}_1-\underline{x})
	+i\underline{k}_2\cdot(\underline{x}_2-\underline{x})
	-i\underline{k}'_1\cdot(\underline{x}_1-\underline{x}')
	-i\underline{k}'_2\cdot(\underline{x}_2-\underline{x}')
	}
	\sum_{\lambda_1\lambda_2}V_2^{\lambda\lambda_1\lambda_2}(V_3^{\lambda'\lambda_1\lambda_2})^*.
\end{equation}
Then, using (\ref{eq:28}) and (\ref{eq:29}), it is straightforward to get
\begin{align}
\label{eq:38}
	I^{\lambda\lambda'}=&\frac{4(2\pi)^6(k_{1+}k_{2+})^2}{t_2^2t_3^2(k_1+k_2)^2}\delta(\underline{x}-\underline{x}')\delta(\underline{x}-z\underline{x}_1-(1-z)\underline{x}_2)e^{-i(\underline{x}_1-\underline{x}_2)^2\frac{1}{2}k_+z(1-z)(\frac	{1}{t_2}-\frac	{1}{t_3})} \nonumber \\
	&\times k_+^2\left\{[z^2+(1-z)^2](\underline{x}_1-\underline{x}_2)^2\delta_{\lambda\lambda'}+2[z(1-z)]^2\underline{\epsilon}^\lambda\cdot(\underline{x}_1-\underline{x}_2)\cdot\underline{\epsilon}^{\lambda'}\cdot(\underline{x}_1-\underline{x}_2)\right\}
\end{align}
where $z=\frac{k_{1+}}{(k_1+k_2)_+}\equiv \frac{k_{1+}}{k_+}$. Write
\begin{equation}
	d^2x_1d^2x_2=d^2x_{12}d^2\tilde{x}
\end{equation}
with $\underline{x}_{12}=\underline{x}_1-\underline{x}_2$ and $\tilde{\underline{x}}=z\underline{x}_1+(1-z)\underline{x}_2$. Using (\ref{eq:27}), and a similar expression for the Fourier transform of $L'$ one finally gets
\begin{align}
\label{eq:40}
		\bar{w}=&\frac{-g^2(N_c^2-1)}{8(2\pi)^4}\int \frac{dt_1dt_2dt_3dt_4}{(t_1t_2t_3t_4)^2}\frac{dk_+}{k_+}k_+^4
		e^{\frac{-i(\underline{x}-\underline{x}_a)^2k_+}{2t_1}+\frac{i(\underline{x}-\underline{x}_b)^2k_+}{2t_4}-i\underline{x}_{12}^2\frac{1}{2}k_+z(1-z)(\frac{1}{t_2}-\frac{1}{t_3})} \nonumber \\
		&\times [z(1-z)]^2\underline{\epsilon}^{\lambda}\cdot(\underline{x}-\underline{x}_a)\underline{\epsilon}^{\lambda'}\cdot(\underline{x}-\underline{x}_b)d^2x d^2 x_{12} x_{12}^2 \left[(\frac{z}{1-z}+\frac{1-z}{z})\delta_{\lambda\lambda'}+2z(1-z)\frac{\underline{\epsilon}^\lambda\cdot \underline{x}_{12}\underline{\epsilon}^{\lambda'}\cdot \underline{x}_{12}}{x_{12}^2}\right].
\end{align}

Now
\begin{equation}
	2d^2x_{12}\frac{\underline{\epsilon}^\lambda\cdot\underline{x}_{12}\underline{\epsilon}^{\lambda'}\cdot\underline{x}_{12}}{x_{12}^2}\rightarrow d^2x_{12}\delta_{\lambda\lambda'}
\end{equation}
since the rest of (\ref{eq:40}) depends only on $\underline{x}_{12}^2$ but not on the orientation of $\underline{x}_{12}$. Now write $t_i=-2/\tau_i$ to get
\begin{align}
\label{eq:42}
	\bar{w}=&\frac{-g^2(N_c^2-1)}{256(2\pi)^4}\int_0^\infty d\tau_4 \int_0^\infty d\tau_1 \int_{\tau_1}^\infty d\tau_3 \int_{\tau_1}^\infty d\tau_2 e^{ i(\underline{x}-\underline{x}_a)^2k+\tau_1/4-i(\underline{x}-\underline{x}_b)^2k_+ \tau_4/4 }d^2xd^2x_{12}x_{12}^2  \nonumber \\
	&e^{ ix_{12}^2k_+z(1-z)(\tau_2-\tau_3)/4 }
	\frac{dk_+}{k_+}dz k_+^4 [z(1-z)]^2(\underline{x}-\underline{x}_a)\cdot(\underline{x}-\underline{x}_b)\left[\left(\frac{1}{z}+\frac{1}{1-z}\right)+\left(-2+z(1-z)\right)\right]. 
\end{align}
In arriving at (\ref{eq:42}) we have taken $d\tau_2d\tau_3\theta(\tau_2-\tau_1)\theta(\tau_3-\tau_2)=\frac{1}{2}d\tau_2d\tau_3\theta(\tau_2-\tau_1)\theta(\tau_3-\tau_1)$, which corresponds to taking the real part of $\bar{w}$. (Taking the graph where the self energy is in the complex conjugate amplitude along with the graph \ref{fig:2b} automatically leads to a real contribution.)

(\ref{eq:42}) has several divergences which are most clearly seen by doing the $d\tau_i$ integrals in (\ref{eq:42}),
\begin{equation}
\label{eq:43}
	\bar{w}=\frac{-g^2(N_c^2-1)}{(2\pi)^4}\int \frac{dk_+}{k_+} \int_0^1 dz d^2x\frac{d^2x_{12}}{x_{12}^2} \frac{(\underline{x}-\underline{x}_a)\cdot(\underline{x}-\underline{x}_b)}{(\underline{x}-\underline{x}_a)^2(\underline{x}-\underline{x}_b)^2}\left[\left(\frac{1}{z}+\frac{1}{1-z}\right)+\left(-2+z(1-z)\right)\right].
\end{equation}
The singularity in $\frac{d^2x_{12}}{x_{12}^2}$ at $x_{12}^2=0$ comes from $\tau_2,\tau_3\rightarrow\infty(t_2,t_3\rightarrow 0)$, and it is an ultraviolet divergence which is cancelled by the graph of figure \ref{fig:3b}. The divergence in $\frac{d^2x_{12}}{x_{12}^2}$ at $x_{12}^2=\infty$ comes from $\tau_2,\tau_3\rightarrow\tau_1$ and it is also an ultraviolet divergence. The $(\frac{1}{z}+\frac{1}{1-z})$ parts of the divergence are cancelled by vertex and fermion self energy corrections (see Appendix A), while the $\int dz(2-z(1-z))=-\frac{11}{12}$ coefficient of the $x_{12}^2\rightarrow\infty$ divergence must be removed by coupling renormalization. It is the only actual divergence encountered at the one loop level.

From the discussion in section \ref{sec:2} it should be clear that the graph of figure \ref{fig:2a}, the decay graph, will be given by (\ref{eq:42}), or (\ref{eq:43}), with the replacements
\begin{equation}
	\underline{x},\underline{x}_a\rightarrow\underline{\theta},\underline{\theta}_a;\ d^2 x,d^2x_{12}\rightarrow d^2\theta,d^2\theta_{12}.
\end{equation}
Here the $\frac{d^2\theta_{12}}{\theta_{12}}$ divergences, as $\theta_{12}^2\rightarrow 0$, is a collinear divergence which is cancelled, when one agrees not to distinguish 2 nearly parallel moving gluons from the parent gluon, by the graph of figure \ref{fig:3a}. The divergence at large $\theta_{12}$ is a genuine ultraviolet divergence which must be removed by renormalization. (Recall that we work in a frame where $\theta_{ab}$ is extremely small so that the ultraviolet divergence here corresponds to $\theta_{12}^2\gg \theta_{ab}^2$.)

Formally, the spacelike-timelike correspondence is exact. In the case of divergences in $\frac{d^2x_{12}}{x_{12}}$ and $\frac{d^2\theta_{12}}{\theta_{12}}$ when $x_{12}^2,\theta_{12}^2 \rightarrow 0$ the correspondence remains exact because there are cancelling divergences between graphs in figure \ref{fig:2} and in figure \ref{fig:3} which eliminate the divergences so that in fact there are no divergences coming from $t_2,t_3\rightarrow \infty$(timelike) or $t_2,t_3\rightarrow 0$(spacelike). On the other hand the divergences in the $(-2+z(1-z))$ part of (\ref{eq:43}) coming from the $t_2,t_3\rightarrow t_1,\underline{x}_{12}^2\rightarrow \infty$ region of (\ref{eq:43}) and from the corresponding $\theta_{12}^2\rightarrow \infty$ part of the timelike graphs are real divergences and must be removed by renormalization, and the renormalization will destroy the correspondence.

It might seem that (\ref{eq:43}) also has a divergence in the $d^2x$ integration at $|x|\rightarrow \infty$, but when all corrections to the $\underline{x}_a$ and $\underline{x}_b$ lines are taken the dipole kernel
\begin{equation}
	\frac{x_{ab}^2}{(\underline{x}-\underline{x}_a)^2(\underline{x}-\underline{x}_b)^2}
\end{equation} 
emerges rather than the
\begin{equation}
	\frac{(\underline{x}-\underline{x}_a)\cdot(\underline{x}-\underline{x}_b)}{(\underline{x}-\underline{x}_a)^2(\underline{x}-\underline{x}_b)^2}
\end{equation}
appearing in (\ref{eq:43}) so that in fact there is no $d^2x$ divergence. 

Let me summarize the various collinear and ultraviolet divergences in the graphs of figure \ref{fig:2} and figure \ref{fig:3}.
\begin{enumerate}
	\item Graph \ref{fig:2a} has a collinear divergence at $t_2,t_3\rightarrow \infty$ corresponding to an ultraviolet divergence in graph \ref{fig:2b} at $t_2,t_3\rightarrow 0$. These divergences are cancelled by corresponding divergences in graphs \ref{fig:3a} and \ref{fig:3b} if one uses graphs \ref{fig:2a} and \ref{fig:3a} in a jet measurement and graphs \ref{fig:2b} and \ref{fig:3b} in a scattering.
	\item Graphs \ref{fig:2a} and \ref{fig:2b} also have ultraviolet divergences coming at $t_2,t_3\rightarrow t_1$ in each case. The $(\frac{1}{z}+\frac{1}{1-z})$ parts of these graphs cancel with other corrections, vertex and quark self energies, around $t_1$. However, the $(-2+z(1-z))$ divergences in these graphs need renormalization which breaks the spacelike-timelike correspondence.
	\item In graph \ref{fig:2a} there is an ultraviolet divergence coming from $(\underline{x}_1-\underline{x}_2)^2\rightarrow 0$ at $t_2,t_3\rightarrow 0$, cancelled by a similar divergence in graph \ref{fig:2b}, and another ultraviolet divergence coming from $(\underline{x}_1-\underline{x}_2)^2\rightarrow \infty$ at $t_2,t_3\rightarrow t_1$. In appendix B we show that both of these divergences correspond to large transverse momentum divergences.
\end{enumerate}

\section{"Conformal" QCD; using the correspondence}
\label{sec:4}
\subsection{"Conformal" QCD}
The fact that field theories with coupling renormalization are immensely more difficult to deal with than theories without coupling renormalization is well illustraded in the classic paper of Gell-Mann and Low\cite{14}. They developed the renormalization group in the context of QED and found it difficult to get explicit results because of coupling renormalization. However, they observed that by dropping all photon self energy graphs the theory became much simpler because the coupling was not renormalized. This conformal QED remains a nontrivial theory, although it does lack unitarity. While in many ways the renormalization of QED and QCD are similar, coupling renormalization in QCD is not associated with a particular set of graphs in any known gauge. In section \ref{sec:3} and in appendix A we have seen that coupling renormalization breaks the spacelike-timelike correspondence between the lightcone wavefunction and decay probabilities. We will explore that breaking a little farther on, but here we focus on a way to define a conformal QCD similar to what was done in QED.

Refer back to (\ref{eq:43}). There are divergences in the $d^2x_{12}$ integration both at $x_{12}^2=0$ and at $x_{12}^2=\infty$. The divergence at $x_{12}^2=0$ in (\ref{eq:43}) corresponding to graph \ref{fig:2b} is cancelled by a similar divergence in graph \ref{fig:3b} as we have already discussed while the divergence at $x_{12}^2=\infty$ cancels, for the $(\frac{1}{z}+\frac{1}{1-z})$ part of (\ref{eq:43}), as we shall see in appendix A. Thus only the $(-2+z(1-z))$ part of the $x_{12}^2=0$ divergence of (\ref{eq:43}) is uncancelled, and this is the coupling renormalization divergence which must be removed in order to evaluate the lightcone wavefunction, and high energy scattering, in QCD.

Instead of renormalizing (\ref{eq:43}), coming from graph \ref{fig:2b}, suppose we just drop the $x_{12}^2\rightarrow \infty$ divergence in the $(-2+z(1-z))$ part of (\ref{eq:43}). Would this give a conformal theory? The problem here is that the $x_{12}^2\rightarrow \infty$ divergence and the $x_{12}^2\rightarrow 0$ divergence of (\ref{eq:43}) cannot be separated without introducing a separation scale and then dropping the $x_{12}^2\rightarrow \infty$ divergence would depend on that scale as would the resulting lightcone wavefunction. However, there is a procedure which does not introduce any scale, and that consists in dropping the $(-2+z(1-z))$ terms in both graph \ref{fig:2b} and in graph \ref{fig:3b}. In terms of singularities we are dropping the $x_{12}^2\rightarrow 0$ singularities of the $(-2+x(1-z))$ parts of these graphs, which cancel in any case, as well as the $x_{12}^2\rightarrow \infty$ singularity of the $(-2+x(1-z))$ part of graph \ref{fig:2b}, the only genuine divergence at this order. This same rule for dropping the $(-2+z(1-z))$ part of self-energy grphs can also be applied to decay graphs, graphs \ref{fig:2a} and \ref{fig:3a} in the current discussion. The remaining graphs, including the $(\frac{1}{z}+\frac{1}{1-z})$ parts of the self-energy contribution will be conformal and obey the spacelike-timelike correspondence. We emphasize, however, that we have demonstrated this correspondence, and the confromality, only at NLO level. At NNLO one must deal with coupling renormalization of a three gluon vertex where the three gluons share longitudinal momentum more or less equally, and that is beyond what has been considered in appendix A.

\subsection{Possible uses of the correspondence}

Our purpose in this paper is to see how the spacelike-timelike correspondence works graphically and pinpoint exactly where and how it breaks down. We have also seen that it is possible to discard a well-defined part of self-energy graphs in order to maintain the conformality and the correspondence . While our purpose is not here to use the correspondence to do calculations it is, perhaps, useful to see how this could come about and to make connection with previous work.

The earliest discussions\cite{5,6,7}, observing that the BFKL equation governs certain observables in jet decay, in addition to determining high energy scattering, did not suggest a general relationship between decay and high energy scattering. That came in the paper of Hofman and Maldacena, in the context of AdS/CFT calculation, and by Hatta\cite{11} who extended the discussion to $N=4$ SYM perturbation theory. Hatta and collaborators\cite{12} extended the discussion to a comparison of resummed kernels in the BK and BMS equations. In all these cases running coupling effects do not enter, or were not considered. The most ambitious attempt to use the spacelike-timelike correspondence has been by Caron-Huot\cite{15} who was able to get all the non $\beta$-function dependent parts of the NLO kernel of Balitsky and Chirili\cite{16} by transforming decay calculations into wavefunction caculations. 

The general problem is that $\beta$-function terms will not obey the correspondence. However, in lightcone gauge the number of self-energy graphs that occur, for example, in the NLO kernel for the BK equation is very small compared to the total sum of graphs. We believe that the full NLO BK kernel could be obtained by doing the corresponding timelike calculation without the $(-2+z(1-z))$ parts of the self-energy graphs, transforming that calculation to the spacelike case and then adding the $(-2+z(1-z))$ parts of the self-energy graphs to the lightcone wavefunction including the renormalization of these self-energy graphs.

Of course Balitsky and Chirilli have already done the NLO calculation of the BK kernel, including the very difficult Fourier transforms to get from momentum to coordinate space which Fourier transforms are not necessary when using the spacelike-timelike correspondence, so there is not so much motivation for doing the calculation the way we suggest in the previous paragraph. However , it may well be that there are other calculations which are more easily done in, say, the timelike case and then transforming to the spacelike case. If such a calculation has many parts it could well be easier to separate out the $(-2+z(1-z))$ parts of the self-energy graphs in the timelike calculation and add them back in the spacelike calculation.

\acknowledgments

The author wishes to thank Y. Hatta, E. Iancu and D. Triantafyllopoulos for stimulating discussions when this work was starting. This work was in part supported by a DOE grant.

\appendix

\section{Appendix A}
The purpose of this appendix is to see that the $\left(\frac{1}{z}+\frac{1}{1-z}\right)$ terms (see for example (\ref{eq:43})) appearing in self-energy graphs cancel with vertex and other self-energy graphs. We limit our discussion to an example and in this example we use a very physical argument rather than a detailed computation. This example should make clear how the cancellation occurs in a more general setting.

\begin{figure}[H]
\begin{subfigure}[b]{1.0\textwidth}
	\centering
	\includegraphics[width=0.5\textwidth]{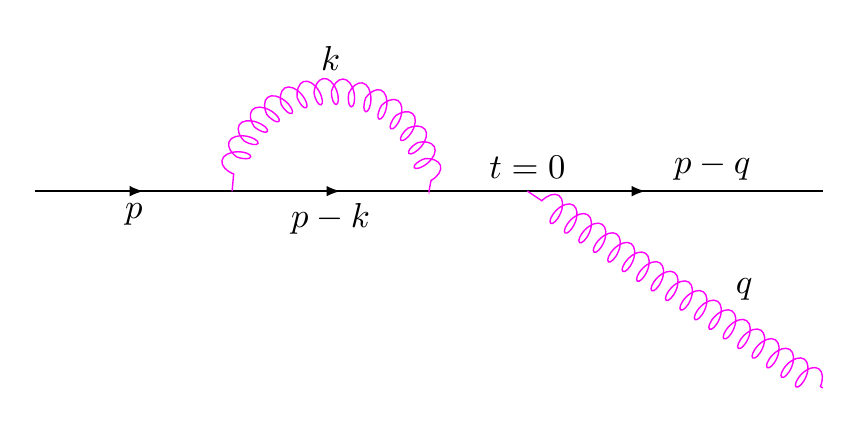}
	\caption{($a$).}\label{fig:4a}
\end{subfigure}

\begin{subfigure}[b]{0.5\textwidth}
	\centering
	\includegraphics[width=1.0\textwidth]{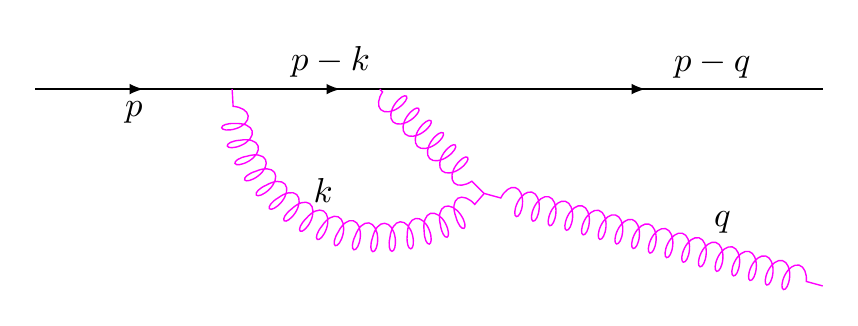}
	\caption{($b_1$).}\label{fig:4b1}
\end{subfigure}
\begin{subfigure}[b]{0.5\textwidth}
	\centering
	\includegraphics[width=1.0\textwidth]{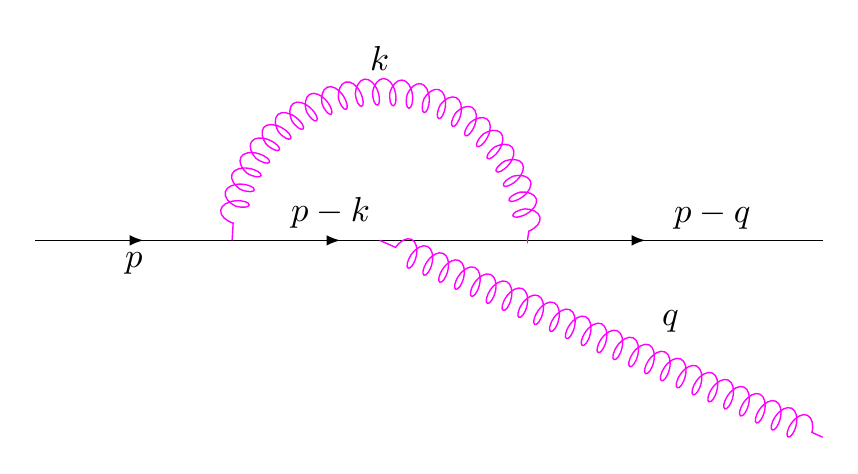}
	\caption{($b_2$).}\label{fig:4b2}
\end{subfigure}

\begin{subfigure}[b]{0.5\textwidth}
	\centering
	\includegraphics[width=1.0\textwidth]{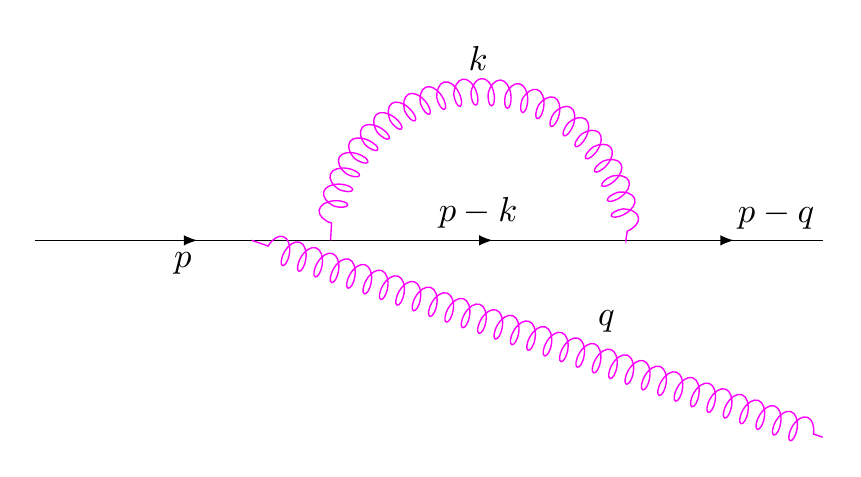}
	\caption{($c_1$).}\label{fig:4c1}
\end{subfigure}
\begin{subfigure}[b]{0.5\textwidth}
	\centering
	\includegraphics[width=1.0\textwidth]{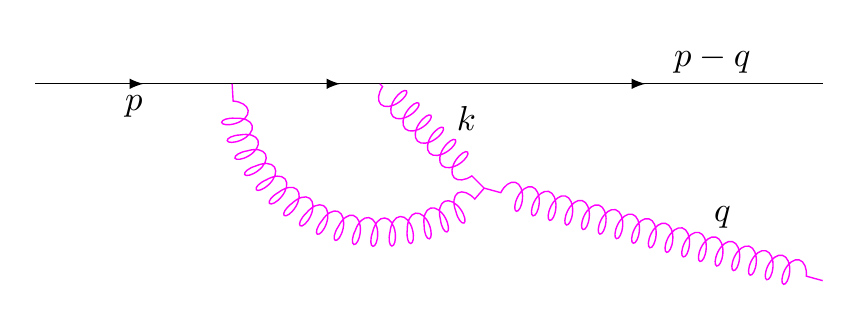}
	\caption{($c_2$).}\label{fig:4c2}
\end{subfigure}

\begin{subfigure}[b]{0.5\textwidth}
	\centering
	\includegraphics[width=1.0\textwidth]{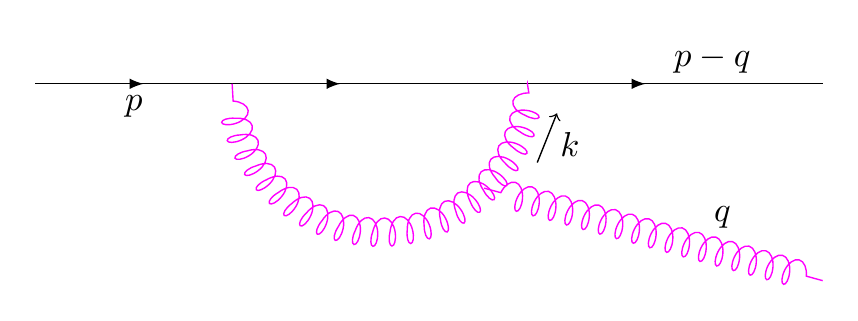}
	\caption{($c_3$).}\label{fig:4c3}
\end{subfigure}
\begin{subfigure}[b]{0.5\textwidth}
	\centering
	\includegraphics[width=1.0\textwidth]{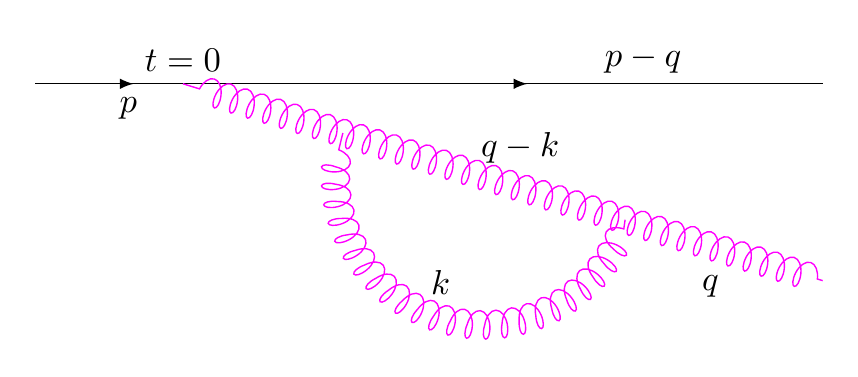}
	\caption{($c_4$).}\label{fig:4c4}
\end{subfigure}

\caption{Running coupling graphs.}\label{fig:4}
\end{figure}

The graphs we analyze are for the renormalization of the quark gluon coupling and are shown in figure \ref{fig:4}, and grouped into $a$, $b$ and $c$ components. Graph \ref{fig:4c4} is the same as appears as parts of the graphs in figure \ref{fig:2}. We always assume that $q_+$ obeys $q_+/p_+\ll 1$. The variable $z$ is given as $k_+/q_+=z$ and in graph \ref{fig:4c4} it is clear that $0<z<1$ and the $z=0$ singularity occurs when the line $k_+\rightarrow 0$ while the $z=1$ singularity occurs when $(q-k)_+\rightarrow 0$ so in each case it is a soft longitudinal momentum. It is, perhaps, clear that such singularities cannot be present in a running coupling renormalization but let's see in detail how the cancellation comes about. The lifetime of the $k$-gluon fluctuation is
\begin{equation}
	\tau_k\simeq \frac{2k_+}{k_\perp^2}
\end{equation}
and $k_\perp^2$ goes large for a divergent term corresponding to a potential coupling renormalization. We take $t=0$ to be the time at the quark, $q$-gluon vertex. Then the maximum time $t$, that $k$ can be emitted or absorbed is $|t|\lesssim \tau_k$. During the time $\tau_k$ the separation of the $q$-gluon from the quark is
\begin{equation}
	\Delta x_\perp\sim \tau_k\cdot \frac{q_\perp}{q_+}=\frac{2}{k_\perp}\cdot z \cdot \frac{q_\perp}{k_\perp} \ll \frac{1}{k_\perp}
\end{equation}
while the transverse wavelength of the $k$-gluon is $1/k_\perp$. Thus the $k$-gluon does not resolve the quark-$q$-gluon pair so that the contribution of $c_1+c_2+c_3+c_4$ is exactly the same as the contribution of $a$ and the sum of $a+c$ is just the negative of the probability that the quark $p$ emit a gluon. $b$ is the probability that quark $p$ emit a soft gluon so $b+(a+c)=0$ by probability conservation.

The argument given above is subtle, however. Let me list values for the divergent parts of the graphs and then comment on why the argument given above does lead to the correct result. The values of the graphs are, taking the divergent quantity $\int\frac{d p_\perp^2}{p_\perp^2}=L$:
\begin{align}
\label{eq:A1}
	(a)&=\frac{-\alpha C_F}{2\pi}L\int_0^\infty\frac{dz}{z} \nonumber \\
	(b_1)+(c_2)&=\frac{+3\alpha N_c}{8\pi}L\int_0^1 dz\left(\frac{1}{z}+\frac{1}{1-z}\right) \nonumber \\
	(b_2)&=\frac{-\alpha}{\pi}(C_F-\frac{N_c}{2})L\int_0^\infty\frac{dz}{z} \nonumber \\
	(c_1)&=\frac{-\alpha C_F}{2\pi}L\int_0^\infty\frac{dz}{z}  \\
	(c_3)&=\frac{+\alpha N_c}{4\pi}L\int_1^\infty dz\left(\frac{1}{z}+\frac{1}{z-1}\right) \nonumber \\
	(c_4)&=\frac{-\alpha N_c}{4\pi}L\int_0^1 dz\left(\frac{1}{z}+\frac{1}{1-z}\right) \nonumber
\end{align}
The $C_F$ terms cancel between $(a)$, $(b_2)$ and $(c_1)$. If we identify the $\int_1^\infty\frac{dz}{z-1}$ integration in $(c_3)$ as equal to $\int_0^\infty\frac{dz}{z}$ the sum of the contributions in (\ref{eq:A1}) vanish. We could also write
\begin{align}
	Z_2-1&=a+c_1, \frac{Z_3-1}{2}=c_4 \nonumber \\
	\frac{1}{Z_1}-1&=b_1+c_2+b_2+c_3
\end{align}
in which case the cancellation is a $Z_1$ cancellation with $Z_2$, $Z_3$.

The subtlety in the above argument is that we take exactly $\left(\frac{1}{z}+\frac{1}{1-z}\right)$ and not $\left(\frac{1}{z}+\frac{1}{1-z}+\mathrm{const.}\right)$ in our expectation of the vertex-self-energy cancellation. The reason for expecting the cancellation in the pole-terms alone is that only graph $c_4$ has other than pole terms. Thus in (\ref{eq:43}) we separate the gluon self-energy terms into pole terms and all the rest. The pole terms cancel as demonstrated above while the remaining $\int_0^1 dz(-2+z(1-z))=-\frac{11}{12}$, the gluonic contribution to the $\beta$-function.

\section{Appendix B}	
\begin{figure}[H]
\begin{subfigure}[b]{1.0\textwidth}
	\centering
	\includegraphics[]{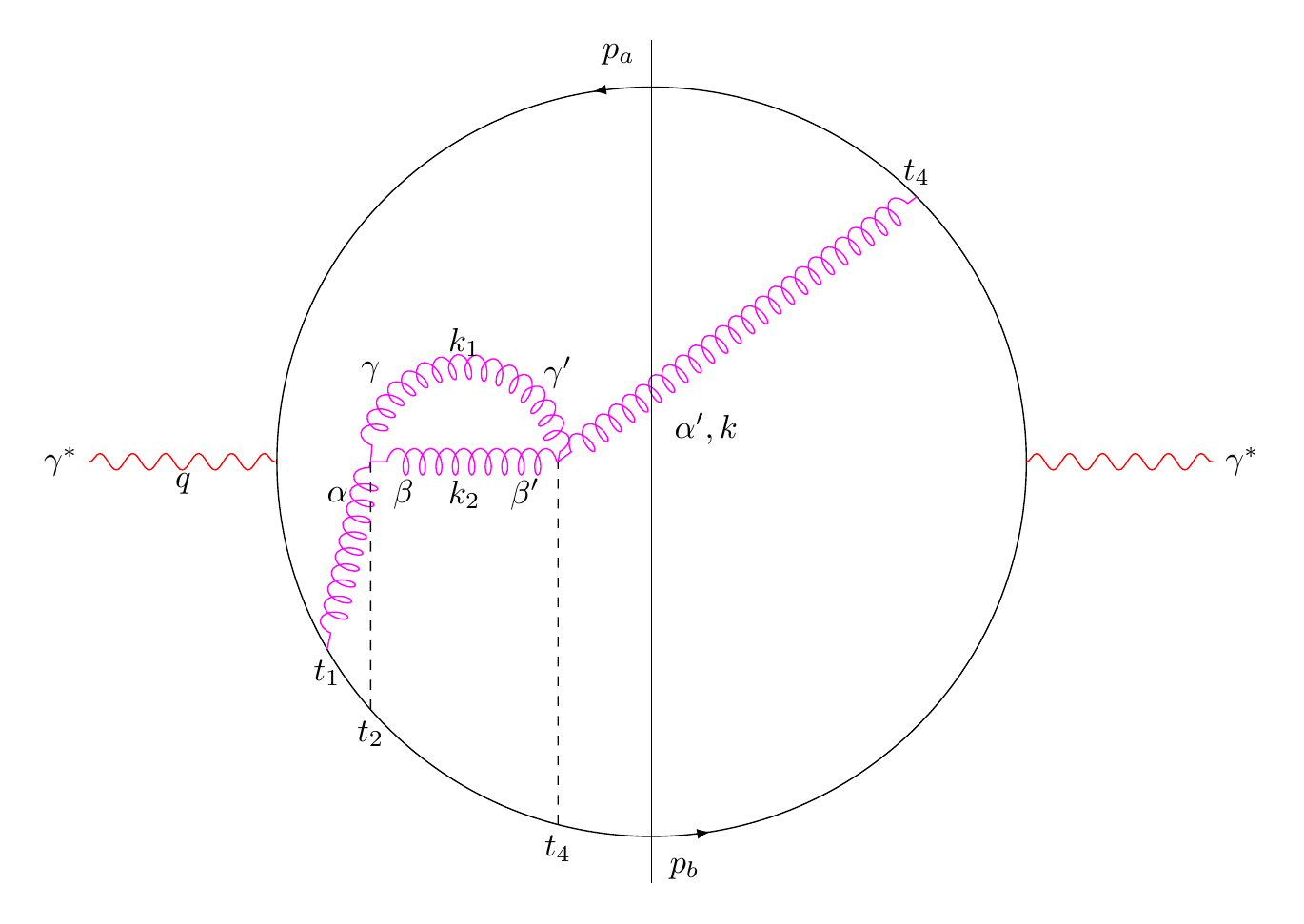}
	\caption{fig:2a.}\label{fig:2a}
\end{subfigure}

\begin{subfigure}[b]{1.0\textwidth}
	\centering
	\includegraphics[]{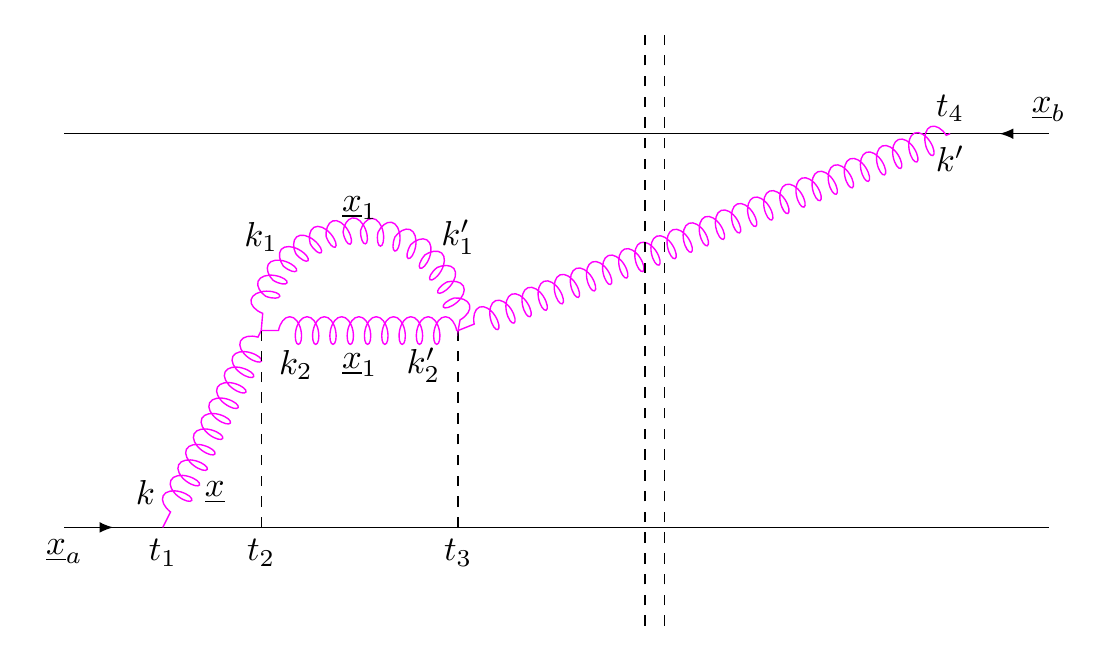}
	\caption{fig:2b.}\label{fig:2b}
\end{subfigure}
\caption{Running coupling graphs.}\label{fig:2}
\end{figure}

\begin{figure}[H]
\begin{subfigure}[b]{1.0\textwidth}
	\centering
	\includegraphics[]{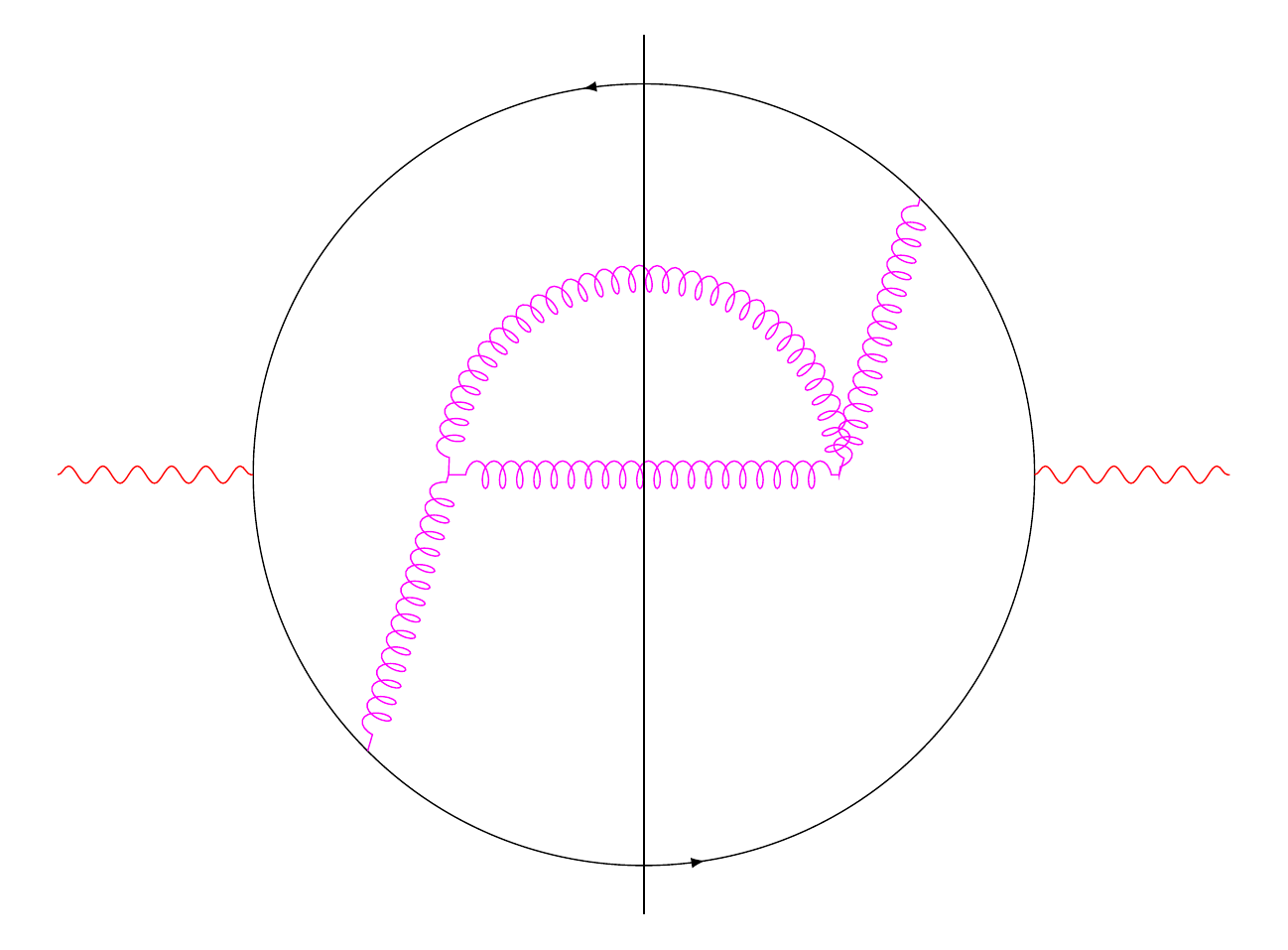}
	\caption{fig:3a.}\label{fig:3a}
\end{subfigure}

\begin{subfigure}[b]{1.0\textwidth}
	\centering
	\includegraphics[]{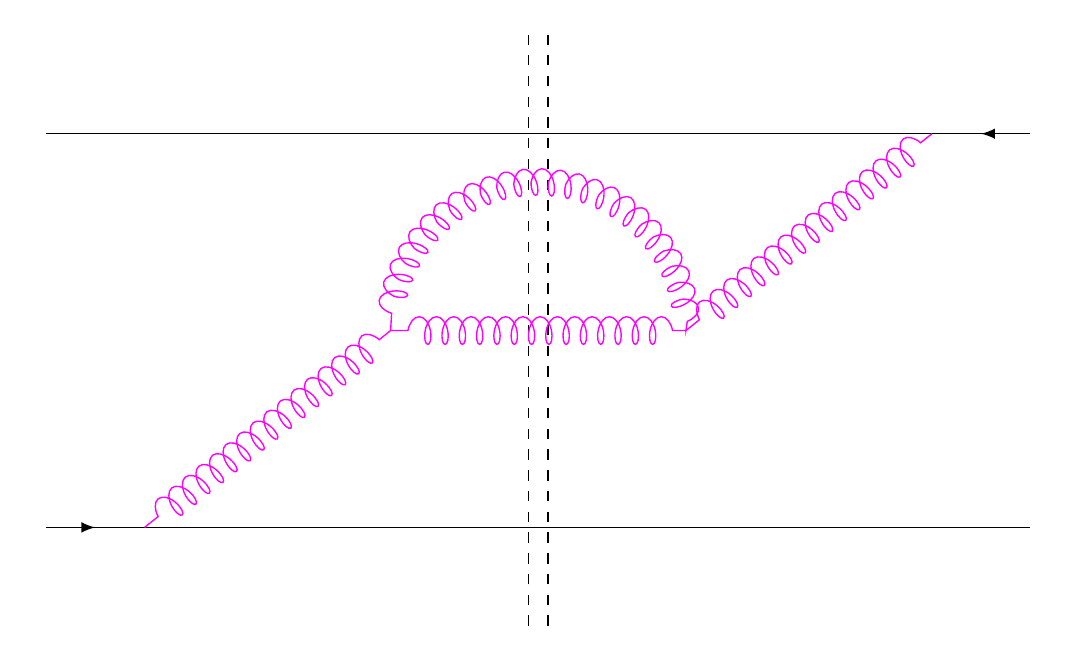}
	\caption{fig:3b.}\label{fig:3b}
\end{subfigure}
\caption{Real graphs corresponding to the graphs of figure \ref{fig:2}.}\label{fig:3}
\end{figure}

In section \ref{sec:3} we have seen that in the correspondence between the graphs in figure \ref{fig:2a} and figure \ref{fig:2b} there are two different ultraviolet divergences which occur in graph \ref{fig:2b}, one with $(\underline{x}_1-\underline{x}_2)^2\rightarrow 0$ and $t_2,t_3\rightarrow 0$ and the other with $(\underline{x}_1-\underline{x}_2)^2\rightarrow \infty$ and $t_2,t_3\rightarrow t_1$. In terms of the correspondence with graph \ref{fig:2a} the $(\underline{x}_1-\underline{x}_2)^2\rightarrow 0$ divergence corresponds to a collinear singularity of graph \ref{fig:2a} while the $(\underline{x}_1-\underline{x}_2)^2\rightarrow \infty$ divergence corresponds to an ultraviolet divergence of graph \ref{fig:2a}. In this appendix we demonstrate that in terms of momentum variables the $(\underline{x}_1-\underline{x}_2)^2\rightarrow 0$ and $(\underline{x}_1-\underline{x}_2)^2\rightarrow \infty$ divergence correspond to $\underline{k}_1^2(\mathrm{or}\ \underline{k}_1^{'2})\rightarrow \infty$ and hence are genuine ultraviolet divergences.

The phase factors centered around the vertex at $t_2$ in graph \ref{fig:2b} are $e^{i\varepsilon}$ where (see (\ref{eq:22})-(\ref{eq:25}))
\begin{equation}
	\varepsilon = \left(\frac{\underline{k}_1^2}{2k_{1+}}+\frac{\underline{k}_2^2}{2k_{2+}}-\frac{(\underline{k}_1+\underline{k}_2)^2}{2k_{+}}\right)t_2+\underline{k}_1\cdot\underline{x}_1+\underline{k}_2\cdot\underline{x}_2-(\underline{k}_1+\underline{k}_2)\cdot\underline{x}.
\end{equation}
Call $(\underline{k}_1+\underline{k}_2)=\underline{p}$, then $d^2k_1d^2k_2=d^2k_1d^2p$ and our object is to see what values of $\underline{k}_1$ and $\underline{k}_2=\underline{p}-\underline{k}$ are dominant in the integration in (\ref{eq:37}) leading to the coordinate space formula (\ref{eq:38}).

One easily finds
\begin{equation}
\label{eq:B2}
	\varepsilon=\frac{t_2}{2k_+z(1-z)}\left[\underline{k}_1-z\underline{p}-(\underline{x}_2-\underline{x}_1)\frac{k_+z(1-z)}{t_2}\right]^2+\underline{p}\cdot\left[-\underline{x}+z\underline{x}_1+(1-z)\underline{x}_2\right]-\frac{(\underline{x}_1-\underline{x}_2)^2}{2t_2}k_+z(1-z).
\end{equation}
\begin{enumerate}
	\item When $(\underline{x}_1-\underline{x}_2)^2$ is small and $t_2\sim (\underline{x}_1-\underline{x}_2)^2k_+$ the first term on the right hand side of (\ref{eq:B2}) is of order
	\begin{equation*}
		\varepsilon\sim (\underline{x}_1-\underline{x}_2)^2\left(\underline{k}_1-z\underline{p}-\mathcal{O}\left(\frac{1}{\sqrt{(\underline{x}_1-\underline{x}_2)^2}}\right)\right)^2
	\end{equation*}
	so that
	\begin{equation*}
		|\underline{k}_1-z\underline{p}|\sim \frac{1}{\sqrt{(\underline{x}_1-\underline{x}_2)^2}}\pm\mathcal{O}\left(\frac{1}{\sqrt{(\underline{x}_1-\underline{x}_2)^2}}\right)
	\end{equation*}
	or since $\underline{p}^2$ will not be large,
	\begin{equation}
	\label{eq:B3}
		\underline{k}_1^2\sim \frac{1}{(\underline{x}_1-\underline{x}_2)^2}
	\end{equation}
	the expected relationship for an ultraviolet divergence.
	\item When $(\underline{x}_1-\underline{x}_2)^2$ is large and $t_2\sim t_1+\mathcal{O}\left(\frac{t_1^2}{(\underline{x}_1-\underline{x}_2)^2k_+}\right)$ the first term on the right hand side of (\ref{eq:B2}) becomes
	\begin{equation}
		\varepsilon\simeq \frac{t_1}{2k_+z(1-z)}\left(\underline{k}_1-z\underline{p}-\frac{(\underline{x}_2-\underline{x}_1)k_+z(1-z)}{t_1}\right)^2
	\end{equation}
	so that
	\begin{equation}
		\underline{k}_1-z\underline{p}\simeq \frac{(\underline{x}_2-\underline{x}_1)k_+z(1-z)}{t_1}\pm \mathcal{O}\left(\sqrt{\frac{2k_+z(1-z)}{t_1}}\right).
	\end{equation}
	$\underline{k}_1^2$ is again very large although the relationship between $\underline{k}_1(\mathrm{or}\ \underline{k}_2)$ and $\underline{x}_2-\underline{x}_1$ is not so familiar as the one given in (\ref{eq:B3}).
\end{enumerate}

%
%

\nocite{*}
\bibliographystyle{JHEP.bst}
\bibliography{jhep}

\end{document}